# MovementVR: An open-source tool for the study of motor control and learning in virtual reality.

**Cristina Rossi[a,b], Rini Varghese[a,b], Amy J Bastian [a,b]**

[a] Department of Neuroscience, The Johns Hopkins University School of Medicine, Baltimore, MD, 21205, 3 USA; [b] Center for Movement Studies, Kennedy Krieger Institute, Baltimore, MD, 21205, USA.

## Abstract

Virtual reality (VR) is increasingly used to enhance the ecological validity of motor control and learning studies by providing immersive, interactive environments with precise motion tracking. However, designing realistic VR-based motor tasks remains complex, requiring advanced programming skills and limiting accessibility in research and clinical settings. MovementVR is an open-source platform designed to address these challenges by enabling the creation of customizable, naturalistic reaching tasks in VR without coding expertise. It integrates physics-based hand-object interactions, real-time hand tracking, and flexible experimental paradigms, including motor adaptation and reinforcement learning. The intuitive graphical user interface (GUI) allows researchers to customize task parameters and paradigm structure. Unlike existing platforms, MovementVR eliminates the need for scripting while supporting extensive customization and preserving ecological validity and realism. In addition to reducing technical barriers, MovementVR lowers financial constraints by being compatible with consumer-grade VR headsets. It is freely available with comprehensive documentation, facilitating broader adoption in movement research and rehabilitation.

## Introduction

Ecological validity is paramount in the study of motor tasks (Bock and Hagemann, 2010; Faria et al., 2023; Ingram and Wolpert, 2011; Mohr et al., 2023). The term refers to how well findings from laboratory settings translate to real-world scenarios, ensuring relevance to everyday life and effective translation to clinical applications (Faria et al., 2023; Mohr et al., 2023; Schmuckler, 2001). While laboratory tasks inherently limit ecological validity compared to real-world movement studies, they are essential as they allow controlled manipulation of task parameters, precise data collection, and safe rehabilitation training. Mimicking real-world conditions in laboratory tasks may improve ecological validity and generalization, and help bridge the gap between experimental rigor with real-world applicability (as reviewed in Bock and Hagemann, 2010; Faria et al., 2023; Ingram and Wolpert, 2011; Levac et al., 2019; Mohr et al., 2023; Schmuckler, 2001).

In recent years, researchers have increasingly turned to virtual reality (VR) to enhance the realism of motor tasks (Arlati et al., 2022; Faria et al., 2023; Ingram and Wolpert, 2011; Levac et al., 2019). VR headsets offer the capability to display immersive and realistic 3D environments, and rapid advancements in VR technology are continuously enhancing the sense of realism and embodiment (Faria et al., 2023; Haar et al., 2021; Ingram and Wolpert, 2011; Perez-Marcos, 2018; Stanney et al., 2020). Key advancements include wireless connectivity, improved immersion – through higher image resolution, frame rates, and optimized calibration of multisensory feedback – and integrated hand tracking (Apăvăloaiei and Achirei, 2023; Radianti et al., 2020; Stanney et al., 2020), which allows users to interact with virtual objects in a natural manner (Ahmad et al., 2019; Buckingham, 2021; Cesanek et al., 2024; Varela-Aldás et al., 2023).

Despite these new technological advancements, the current uses of VR for behavioral research and rehabilitation remain limited in terms of environmental realism, body representation, and task demands. For example, the tasks are often abstract and not representative of real-world environments and dynamics (as reviewed in Cesanek et al., 2024; Corti et al., 2022; Demers et al., 2021; Haar et al., 2021; Levac et al., 2019). Likewise, task environments are minimalistic (e.g., a black background), task objects are basic geometric shapes (e.g., cubes or spheres), and movements are constrained to limited joints or dimensions (e.g., 2D) (Anglin et al., 2017b, 2017a; Juliano et al., 2022; Wähnert and Gerhards, 2022; Wähnert and Schäfer, 2024). Moreover, participants typically use controllers that dictate object movement in a rigid, pre-programmed way (e.g., objects directly following the controller's position), rather than manipulating objects in a natural way as would be possible with advanced hand-tracking technology. In turn, this setup restricts the natural multiplicity of tasks solutions, as task success can only be achieved with a limited range of movement trajectories or limb poses. These designs limit ecological validity and generalization, consequently, the potential advantages of VR for the study of motor control and learning (Bock and Hagemann, 2010; Corti et al., 2022; Faria et al., 2023; Haar et al., 2021; Ingram and Wolpert, 2011; Levac et al., 2019; Schmuckler, 2001). This may explain why, despite the technological advancements and growing interest in the tool, VR-based tasks for behavioral research and rehabilitation remain less prevalent than traditional experimental or clinical interventions (Corti et al., 2022; Glegg and Levac, 2018; Haar et al., 2021; Levac et al., 2019; Ribeiro-Papa et al., 2016).

A significant obstacle in designing realistic virtual environments and adopting new VR tools for movement research and rehabilitation is the complexity of implementation (Glegg and Levac, 2018; Haar et al., 2021; Kourtesis et al., 2020; Levac et al., 2019; Liu and Chen, 2023; Maymon et al., 2023). Creating realistic VR environments necessitates advanced expertise in computer programming, including object-oriented coding and graphic design (Glegg and Levac, 2018; Jung et al., 2021; Kourtesis et al., 2020; Liu and Chen, 2023; Maymon et al., 2023; Wang et al., 2020; Xie et al., 2019). Concurrently, researchers who design motor tasks must possess advanced expertise in movement neuroscience or a clinical background to ensure that the tasks study the appropriate parameters. Thus, designing realistic motor tasks in VR requires a complex interdisciplinary skillset, posing a barrier to adoption in clinical or academic research settings (Glegg and Levac, 2018; Haar et al., 2021; Levac et al., 2019; Maymon et al., 2023; Sokołowska, 2023). In particular, the computational expertise barrier – and financial burden of alternatively hiring external resources – can make it challenging to develop naturalistic VR tasks for basic and translational academic research. This scarcity of evidence-based VR rehabilitation research may in turn limit the efficacy of commercial VR rehabilitation products. Simplifying the design process to minimize these implementation barriers would enhance access and improve the ecological validity of movement neuroscience research.

In response to this challenge, we have developed the "MovementVR" platform, which simplifies the design of motor tasks in VR by eliminating the need for computer programming expertise. MovementVR is an open-source software tool that allows researchers to design realistic upper extremities tasks performed in VR headsets, facilitating the study of motor performance and motor learning, including adaptive and reinforcement learning. The platform features a realistic 3D environment with naturalistic hand-object interactions, enabled by state-of-the-art hand-tracking technology. The VR components are fully implemented, so researchers do not need to access or modify backend code or graphic design, although these options remain available if desired. The task is highly customizable through a graphical user interface (GUI), enabling researchers to design their paradigms in an intuitive, code-free manner.

The purpose of this paper is to describe the MovementVR platform and applications. In the following sections, we showcase the features of our platform ("Features of MovementVR"), comparing it to existing tools. We then describe the task and platform in detail ("Description of the MovementVR reaching task" and "Description of MovementVR").

**Gaps in existing platforms for design of VR motor tasks**

Several platforms exist to help researchers develop experiments without extensive technical knowledge. However, current platforms are insufficient to implement several aspects of the task presented here. In particular, most tools are not suited for VR experiments in head-mounted displays (HMDs). The few tools targeted at HMDs do not typically provide the hand tracking functionalities needed to achieve realistic hand-object interactions and motor adaptation. We summarize existing tools below and in Figure 1, categorized by the interface they offer.

<u>GUI-based tools.</u> *PsychoPy* (Peirce et al., 2019; Peirce, 2009, 2007) provides a user-friendly Builder interface for basic tasks but requires Python scripting for motor adaptation and gesture-based hand control. In gesture-based hand control, select hand gestures are recognized and mapped to specific actions or controls within the application. This contrasts physics-based object manipulation which replicates intuitive and natural control. For example, in gesture-based hand control, one would pinch to grasp a virtual cup, but in physics-based hand control one would open their hand and then grasp the cup like in the natural world. Similarly, *VRception* (Gruenefeld et al., 2022) provides the "WYSIWYG" ("what you see is what you get") GUI to design the 3D environment directly in the HMD, but any functional behavior (from trial progression to hand tracking) must be implemented in Unity. Even with scripting, both platforms lack built-in features for physics-based object manipulation and support for newer HMDs (*PsychoPy* does not support newer HMD and development via PsychXR has ceased as of June 2024[1], and *VRception* repository has not been updated since January 2022[2]).

---

[1] https://discourse.psychopy.org/t/running-psychxr-on-meta-quest-2/40551
[2] https://github.com/UweGruenefeld/VRception

*NeuroVR* and *NeuroVirtual 3D* GUIs offered limited, gesture-based hand tracking but have been discontinued (Algeri et al., 2009; Cipresso et al., 2016, 2014, Riva et al., 2011, 2007a, 2007b). Other GUI-based platforms are either task-specific – e.g., *OpenVibe* for BCI (Renard et al., 2010), *VREX* for visual cognition (Vasser et al., 2017), and *MazeMaster* for navigation (Bexter and Kampa, 2020) – or commercial – e.g., *Unity Visual Scripting, Virtual Maker, CoSpaces,* and *SimLab VR Studio* – and do not provide a free solution for hand tracking and motor adaptation.

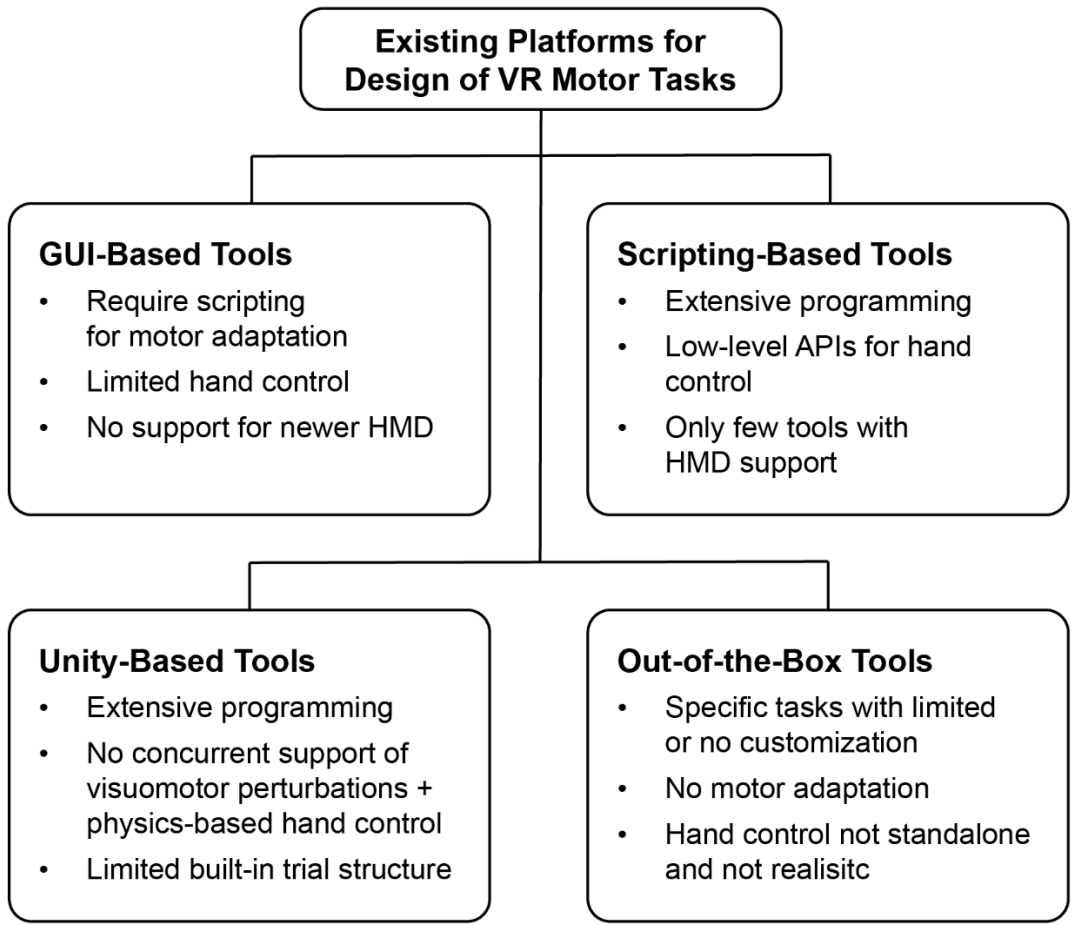

**Figure 1. Existing Platforms for Design of VR Motor Tasks.**

Scripting-based tools. *Ouvrai* (JavaScript) (Cesanek et al., 2024) and *Psychtoolbox* (MATLAB) (Brainard, 1997; Kleiner et al., 2007) can be used to implement tasks like ours but with extensive programming. These platforms rely on low-level APIs (application package interface) for hand tracking (WebXR, Three.js, and OpenXR), so users would need to develop their own hand models and interfaces for robust object manipulation and visuomotor perturbations.

Other platforms like *VR Juggler* (C++/Java) (Bierbaum et al., 2001), *vexptoolbox* (Python) (Schuetz et al., 2023), *PTVR* (Python) (Castet et al., 2024), and *R2VR* (R) (Vercelloni et al., 2021) lack built-in hand tracking functionalities. Unfortunately, most scripting-based platforms for experiment development lack built-in support for HMDs – e.g., *OpenSesame* (Mathôt et al., 2012), *jsPsych* (de Leeuw, 2015; de Leeuw et al., 2023), *Expyriment* (Krause and Lindemann, 2014), *lab.js* (Henninger et al., 2022), *PoMLab* (Takiyama and Shinya, 2016), *OnPoint* (Tsay et al., 2018), *Empirica* (Almaatouq et al., 2021).

Unity-based tools. *HaRT (The Virtual Reality Hand Redirection Toolkit)* (Zenner et al., 2021) uniquely offers built-in tools for hand tracking that include application of visuomotor perturbations, but lacks physics-based object manipulation, and only supports Leap Motion sensors. On the other hand, *HTPK (Hand Physics Toolkit)* (Muresan et al., 2023; Nasim and Kim, 2018), *CLAP* (Verschaar et al., 2018), *RUIS (Reality-based User Interface System)* (Takala, 2014), and *QuickVR* (Oliva et al., 2022) offer varying degrees of built-in support for hand tracking with physics-based object manipulation, but not for visuomotor perturbations. Moreover, *Ubiq-exp* (Steed et al., 2022) and *Rehab-Immersive* (Herrera et al., 2023) focus on gesture-based hand tracking without visuomotor perturbations. All these tools lack the built-in sequential trial structure needed for motor adaptation paradigms, and this is difficult to implement in object-oriented environments like Unity, especially for behavioral researchers used to top-down scripting approaches.

Other Unity tools support paradigm design in varying capacities (sequential trial structure, experimental conditions…), including *UXF* (*Unity Experiment Framework*) (Brookes et al., 2020), *bmlTUX* (*Biomotion Lab Toolkit for Unity Experiments*) (Bebko and Troje, 2020), *Multipurpose Virtual Reality Environment for Biomedical and Health Applications* (Torner et al., 2019), *VRSTK (Virtual Reality Scientific Toolkit)* (Wolfel et al., 2021), *VR-Rides* (Wang et al., 2020), *RemoteLab* (Lee et al., 2022), and *Eve (Experiments in Virtual Environments)* (Grübel et al., 2017). However, these tools lack built-in hand-tracking functionalities, and it would be technically challenging to combine the multiple Unity tools needed to support all aspects of our task.

Out-of-the-box tools. The *NeuroRehabLab* has developed out-of-the-box applications that target motor rehabilitation through engaging games like rowing and gliding, and include bimanual movements and hand tracking via the Leap Motion sensor (e.g., *Neurorehabilitation Training Toolkit*, *RehabNet*, *NeuRow*) (Bermúdez I Badia and Cameirão, 2012; Cameirao et al., 2010; Vourvopoulos et al., 2016, 2014, 2013). Other open-source applications that focus specific tasks relevant to movement research or rehabilitation include *Whack-A-Mole* (pointing task for stroke rehabilitation) (Hougaard et al., 2022), *Planet Juggle* (juggling task) (Adolf et al., 2019), and *SkyBXF* (research modification of the videogame *The Elder Scrolls V: Skyrim VR*) (Liu and Chen, 2023). These applications differ substantially from our task and do not support motor adaptation, standalone hand tracking, realistic object manipulation, or flexible customization.

There are a limited number of published studies that resemble ours in some aspects, but not all, and the associated applications have not been made openly available. Aspects include realistic object manipulation for rehabilitation (Avola et al., 2019; Fernández-González et al., 2019; Ma et al., 2007; Oña et al., 2020; Pereira et al., 2020), bimanual motor adaptation for rehabilitation (Shum et al., 2020), other types of motor adaptation within a realistic environment (Nardi et al., 2023; Wilf et al., 2023).

## **MovementVR offers a affordable, accessible, and versatile solution to realistic motor tasks.**

**Task overview**

MovementVR is a novel platform designed to facilitate the development and execution of realistic motor tasks in virtual reality (VR) with minimal design effort and no need for coding expertise.

The tasks designed with MovementVR are lifting tasks fundamentally based on well documented reaching and transport literature but in a more functional context. On each trial, participants lift an object to a designated target – in the sample task provided, they lift a plate of grapes up to a target with a bird perched on it (Figure 2). Participants' hands are tracked by the app in real-time and represented virtually in a realistic manner: the virtual hand is fully articulated, i.e., each finger joint moves according to the user's real hand, and it interacts with the virtual environment in a physics-based manner, i.e., objects respond to touch as they would in real life. This way, participants see their own hands in the task and use them to lift the virtual object in a way that feels natural.

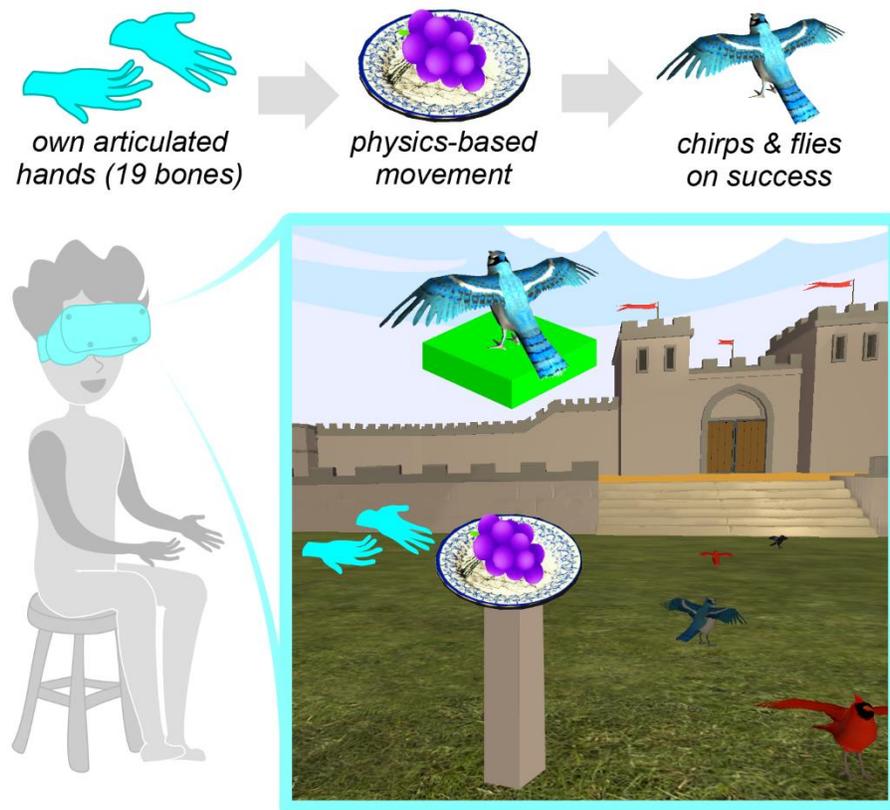

**Figure 2. Schematic of the MovementVR task.**

The task can be modified to study kinematics of arm and hand movement under varied movement contexts, e.g., unimanual or bimanual, different motor learning signals, e.g., error-based adaptation or reinforcement paradigms, as well as perceptual illusions, e.g., size-weight illusions, at the experimenter's discretion. Task parameters and experimental paradigms are highly customizable via an intuitive graphical user interface (GUI) without coding involved. Researchers can run the app on compatible VR headsets (at the time of publication, the Meta Quest 2 and Meta Rift were tested; a regularly updated list of compatible devices is available on the website). Kinematic data is automatically collected by the app and can be analyzed using a provided GUI. We explain each feature in detail in later sections.

The primary aim of MovementVR is to reduce existing barriers that hinder the use of realistic VR tasks for studying reaching and motor learning. To achieve this, we focused on two key aspects: (1) enhancing the realism and relevance of motor tasks to real-life movements, and (2) making the platform accessible for widespread use by researchers and clinicians with diverse backgrounds. This approach resulted in a set of features for the motor task and the platform that are illustrated in Figure 3, and that we discuss below.

**Features of the motor task that enhance realism**

Unconstrained movement. The task permits full degrees of freedom in arm movement, allowing participants to move freely in 3D space as they would in daily life. This addresses the limited degrees of freedom in traditional tasks – particularly those not in VR, such as those using tablets or robots – which typically constrain the workspace to 2D, restrict movement to specific joints, or limit the range of motion (Haar et al., 2021; Ingram and Wolpert, 2011; Levac et al., 2019).

Realistic environment. The VR environment is immersive and realistic, including the landscape and the virtual objects representing everyday items – such as plates, birds, and trees. This addresses the artificial environments of traditional tasks, which often use simple shapes like circles or spheres and minimalistic backgrounds (Haar et al., 2021; Ingram and Wolpert, 2011; Levac et al., 2019).

Natural hands. We use the hand tracking technology integrated into the headset device to track participants' hands in real time. We represent participants' hands realistically within the VR environment, and participants use their own articulated hands to interact with virtual objects. This addresses the unnatural and abstract effectors used in traditional tasks, including those in VR, where participants use controllers that are mapped to single-point effectors within the task (Ahmad et al., 2019; Apăvăloaiei and Achirei, 2023; Buckingham, 2021; Levac et al., 2019).

Lifelike object manipulation and movement: Hand interaction with virtual objects occurs realistically, simulating real-world physics. Objects move and react to hand contact in a manner consistent with physical laws, providing lifelike responses. This addresses the unnatural object dynamics of traditional tasks, including those in VR, which often hard-code object motion so that it responds unnaturally to hand movements – e.g., the object may directly track the position of a hand landmark (Ahmad et al., 2019; Buckingham, 2021; Ingram and Wolpert, 2011; Levac et al., 2019).

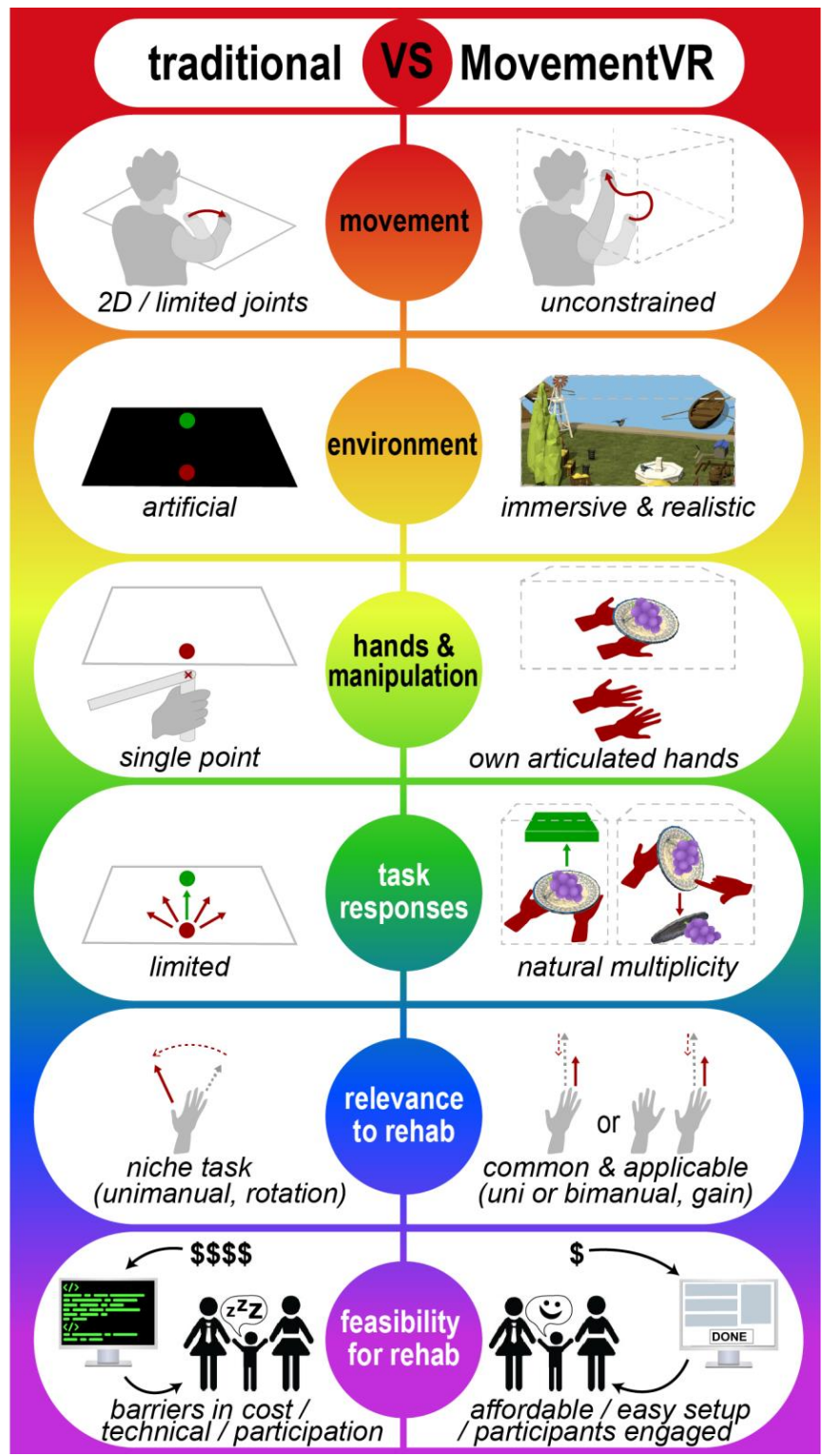

**Figure 3. Features of MovementVR that enhance realism and accessibility (right column) compared to traditional paradigms (left column).**

Natural redundancy of task responses. The combination of realistic hands and object movement allows for an infinite number of task responses and solutions, akin to real-world scenarios (Buckingham, 2021; Levac et al., 2019; Levin et al., 2015; Levin and Demers, 2021). Task success can be achieved through different trajectories or configurations of the virtual object. In turn, each configuration of the virtual object can be achieved through different combinations of hand and finger postures, and also depends on the history of hand-object interactions. This addresses the constrained response variability in traditional tasks, which limit the degrees of freedom of object/cursor motion or base success criteria on a single variable, such as movement angle (Levac et al., 2019; Levin et al., 2015; Levin and Demers, 2021)

Focus on everyday/practical tasks. We select a task that resembles common everyday activities, where most hand movements are bimanual and involve object manipulation (Ingram and Wolpert, 2011; Kilbreath and Heard, 2005). Participants lift an object to a target position – in the sample experiment, they lift a plate containing grapes. This task mimics real-life actions like serving dinner or storing dishes. This addresses the focus of traditional tasks on niche movements, such as reaching in a straight line at a specific angle activities (Haar et al., 2021; Ingram and Wolpert, 2011; Muratori et al., 2013; Ranganathan et al., 2021). Additionally, the platform supports both unimanual and bimanual control. Importantly, bimanual movements are still understudied in motor learning, despite their prevalence in daily activities (Kilbreath and Heard, 2005; Schoenfeld et al., 2021).

Rehabilitation-relevant learning and engagement. The platform includes a gain perturbation feature to train participants to reach higher with one hand than the other. This asymmetry mimics conditions like cortical stroke or cerebral palsy that cause unilateral weakness (Hendricks et al., 2002; Herard et al., 2024; Lodha et al., 2012). Training with perturbations that augment motor error has proven effective for rehabilitation in some patients (Abdollahi et al., 2018; Israely and Carmeli, 2016; Reisman et al., 2013; Shum et al., 2019). Therefore, our task may have applications for physical rehabilitation. This addresses the artificial nature of perturbations studied by traditional adaptation paradigms, such as rotations imposed on a single limb (Babic et al., 2016; Carius et al., 2024; Heald et al., 2023). The realism and immersion of the motor task make it more engaging than traditional paradigms with artificial graphics. This factor further supports the clinical relevance of our platform, as it may promote patients' engagement and motivation, and hence adherence to the rehabilitation treatment (Ambros-Antemate et al., 2023; Faria et al., 2023; O'Brien, 2007; Teo et al., 2022; Zanatta et al., 2022).

**Features of the platform that enhance accessibility.**

Affordability. The app is standalone, requiring only a VR headset with built-in cameras for hand tracking (e.g., Meta Quest 2), costing less than $500. This setup runs the experiment and collects data without needing a computer or external tracking devices. For those with older headsets, an alternative option using other headsets and Leap Motion sensors with a computer is provided. The GUI for experiment design and data analysis is freely available online, offering a cost-effective solution compared to expensive laboratory equipment (Arlati et al., 2022; Levac et al., 2019).

Ease of use and versatile design. The platform facilitates experiment design, execution, and data analysis without computer programming or extensive technical expertise. The intuitive GUIs come with detailed manuals, and the app is easy to install with step-by-step instructions. Experimenters can get started within hours, contrasting with the months needed to code a VR task from scratch. The platform offers highly customizable task parameters (over 500 options), allowing experimenters to design their desired tasks. This versatility does not compromise ease of use, as users are guided to set common parameters, while additional parameters can be accessed through expandable windows. The platform supports the study of motor learning by incorporating both motor adaptation and reinforcement learning. This makes our platform relevant not only for research but also for clinical applications, as motor learning paradigms in VR hold promise for motor rehabilitation (Faria et al., 2023; Haar et al., 2021; Kim et al., 2020; Levac et al., 2019; Levin et al., 2015; Levin and Demers, 2021; Yanovich and Ronen, 2015).

In sum, MovementVR provides an affordable, accessible, and versatile solution for designing VR motor learning experiments without coding expertise. It runs independently on a low-cost VR headset, eliminating the need for expensive lab equipment or external tracking devices. The user-friendly interface allows researchers to quickly design and customize experiments without programming, while still offering powerful options for studying both motor adaptation and reinforcement learning. Its versatility makes MovementVR useful across research and clinical settings, helping advance ecologically valid studies of human movement and rehabilitation. Our platform facilitates the creation of studies that can address these open questions, advancing the understanding and rehabilitation of human movement through ecologically valid research.

## Description of the MovementVR lifting task

### Task elements

The motor task includes the following elements (Figure 4A):

- Plate: the object that participants must lift. On each trial, the plate appears in its starting position on the plate stand. A grape object rests on the plate.
- Target: the goal location that participants must bring the plate to. The target position and size determine the success criteria for the task (success criteria are discussed in detail below). The mesh is the visible shape of a virtual object. For example, the "target mesh" is the visible shape of the target – when turned off, the target is invisible. The target mesh can change color to provide feedback on success or failure.
- Hand home: the home position that hands must return to for the trial to start (see phases below). A hand is considered in the home position if 1) the wrist is within a certain home position volume and 2) all hand bones are outside of a "no-hand zone" (typically, encompassing the space where the plate will appear). An associated "hand home mesh" is responsible for the visual representation of the hand homes and may change in appearance based on whether the hand(s) are in their home positions.

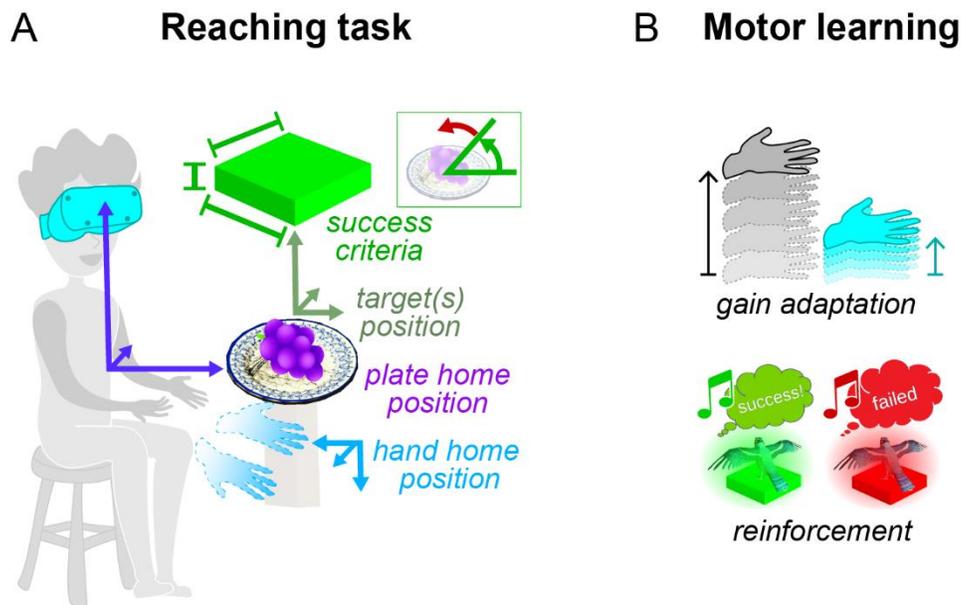

**Figure 4. Annotated schematic of the reaching task elements (A) and motor learning signals (B).**

- <u>Bird:</u> an animated bird object rests on or in the vicinity of the target. Different bird species are available and may rotate. The species rotation as well as bird animations and sounds may change based on the feedback, as discussed below. A visual representation of a bird perch can also be used.
- <u>Feedback message</u>: a cartoon displaying a message related to the task outcome. Both appearance and message change depending on the outcome.
- <u>Hands</u>: a realistic representation of the participants' hands which are tracked in real-time. Each hand has 19 bones that can be used to manipulate the plate object in a realistic manner.
- <u>Workspace</u>: boundaries within which the plate must remain. If the plate exits the task workspace or hits the boundary detecting a fall (typically, the ground), then the trial ends and the outcome is a failure.
- <u>Trial number:</u> text displaying trial number.
- <u>Instruction messages:</u> text instructing participant on what to do in each phase of the trial – the appearance and text change for each phase. Typically displayed on the first trial.
- <u>Background environment:</u> the surrounding landscape and objects unrelated to the task. It includes a medieval garden and port with several birds flying around.

**Trial phases**

Each trial in the motor task consists of the following phases:

1. <u>Intertrial</u>: time delay between trials. The phase ends after a predetermined amount of time.
2. <u>Home position</u>: participants are asked to reposition their hand in the home position. The phase ends after the hands are in the home position for a predetermined amount of time.
3. <u>Plate contact:</u> participants are asked to begin lifting the plate. The phase ends when participants first make contact with the plate.
4. <u>Plate lift:</u> participants are actively lifting the plate and attempting to bring it to the target. The phase ends when the plate reaches the target, or if it falls or exits the task workspace.
5. <u>Feedback:</u> participants receive feedback on whether the trial was a success or a failure (the experimenter chooses if and what type of feedback to give – see MovementVR Builder below). The phase ends after a predetermined amount of time, which may vary depending on the task outcome and associated feedback.

6. <u>Break</u>: participants take a resting break. For trials after which a break is scheduled, the phase ends after a predetermined amount of time. Otherwise, the phase ends immediately (no break).

Additionally, all phases as well as the overall trial have time limits associated with them – in addition to the conditions above, the phase or entire trial ends if this time limit is reached.

**Success and failure criteria**

A trial is successful if all these conditions are met:

a) The plate center reaches the target zone; defined as a region centered around the target position and with size equal to target size
b) The tilt of the plate is smaller than the maximum threshold allowed
c) Conditions a) and b) remain true for a predetermined amount of time

The trial fails if any of the following occur:

- The plate center reaches the target zone, but either conditions a) or b) above become unmet before the predetermined amount of time has passed (the plate exits the target zone and/or the plate tilt is/becomes too large)
- The plate falls or exists the workspace, as described above in "Workspace"
- The time limits for individual phases or for the entire trial are reached, as described above.

**Motor learning applications of MovementVR**

MovementVR supports motor adaptation and reinforcement learning (Figure 4B). Motor adaptation is achieved as follows:

- <u>Gain perturbation.</u> The perturbation consists of a gain that alters the movement of the virtual hand relative to the real hand. There is a "zero position" where the real and virtual hands always align. As the real hand moves away from the zero position, the virtual hand moves in the same direction but at a speed scaled by the gain magnitude. Thus, the distance of the virtual hand from the home position is proportional to that of the real hand, defined by the gain magnitude.

- Adaptation paradigm. The perturbation varies across trials based on the paradigm design. It is zero during baseline and washout phases, it gradually increases from zero to its peak magnitude during the gradual adaptation phase, and it remains at its peak magnitude during the full perturbation phase.
- Backend computations. The task includes computations that 1) select an appropriate "zero position" based on baseline trials and 2) ensure smooth transitions of hand position during abrupt changes in the perturbation gain.

Reinforcement learning is achieved as follows:

- Real-time feedback. The color of the target mesh changes in real-time based on whether the success criteria of plate position and tilt are currently met.
- Feedback phase. During the feedback phase, the following items change depending on task outcome, providing feedback: 1) the color of particle effects around the target and plate, 2) the bird sounds and species swapping, and 3) the feedback messages.

**MovementVR Software Implementation**

**Overview**

MovementVR is an open-source software platform designed to facilitate the creation of reaching tasks in VR without the need for computer programming. It includes three main components available on the website:

MovementVR Builder. The MovementVR Application is highly customizable through the MovementVR Builder, a GUI that allows modification of task parameters and features. The GUI produces input files that are used by the VR application during the experiment.

MovementVR Application. This is the primary implementation for the VR environment and reaching task. Users can download the apk installer (Android Package Kit, similar to an executable file for PC applications), and then install and run this application on the VR headset. For those wishing to access and edit the backend application, the Unity package from which the apk was created will be made available.

MovementVR Analyzer. The MovementVR Application automatically records kinematic data from the hands and other relevant task variables. This data can be analyzed using the MovementVR

Analyzer, a GUI that outputs raw data in a standard format (matching established motion tracking software).

Given these components, the steps to design and run an experiment using MovementVR and analyze the collected data are as follows:

1. Design the paradigm using MovementVR Builder

2. Run the experiment by installing and executing the MovementVR Application on the VR headset device, using input files from step 1.

3. Analyze the recorded data using the MovementVR Analyzer

Each component and the corresponding steps are detailed below.

**Step 1 with MovementVR Builder**

The MovementVR Builder GUI is a point-and-click interface to design the experimental paradigm. The main interface (Figure 5A) provides a graphical representation of a single trial of the motor task, including the "*Task elements"* and the sequence of "*Trial phases"* described in the previous section. Users can customize parameters related to each object or item by clicking on it, which opens a pop-up window called "Static Builder" with editable input fields for the relevant parameters (Figure 5B). Within each pop-up window, related parameters are further grouped into tabs, and expandable windows are used for parameters that are rarely changed (identifiable by their triangle indicator and underlined clickable title).

The main interface also has buttons to modify parameters not associated with specific objects – for example, the "edit transitions" and "edit instructions" buttons allow altering the timing parameters that mediate phase transitions and the instructions shown to the users. "Quick Links" buttons provide quick access to commonly edited parameters.

The interface contains intuitive explanations and illustrations for the different task parameters. Additionally, a Help button at the bottom of the window provides general guidance on using the Static Builder, and "?" icons next to each parameter group provide detailed explanations and instructions for setting parameter values (e.g., coordinate system).

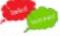

Figure 5. MovementVR Builder main interface (A) and Static Builder pop-up for "target" object (B).

MovementVR Builder allows the experimenter to customize over 500 features of the task and paradigm. Broadly, these include:

- Reaching task parameters, such as starting position of the plate, target position and success criteria, home position for the hands.
- Motor adaptation parameters, such as size and schedule of the gain perturbation, and which hand(s) are perturbed.
- Reinforcement learning parameters, such as if and what feedback to deliver on success or failure trials. For example, outcome-dependent text and appearance of feedback messages, color of the target mesh and particle effects, bird sounds and any outcome-dependent bird swapping rule.
- Trial phases parameters, such as the predetermined time amounts used for transitioning between trial phases
- Tracking and recording selection, such as which hand(s) are used for the task (unimanual or bimanual) and which kinematic data should be recorded
- Appearance and physics properties of plate and grape, such as size and friction coefficients.
- Appearance of target, such as mesh shape and size
- Definition and appearance of home position for hands, such as size and shape of the home position conditions for the wrist and for all bones in the hand, and shape, size, and color feedback of the home position meshes
- Bird appearance and swapping rules, such as order of bird species and when to switch, and bird size, position, selection of animations and sounds
- Definition of workspace, such as shape, size and location, and boundaries to detect falls
- Instructional messages, such as the text to be displayed to guide participants through the trial
- Other objects, like bird perch, plate stand, trial number…

The Static Builder pop-up window allows users to define parameters that remain constant throughout the experiment (with a single editable input field for each parameter). MovementVR also allows users to define parameters that vary across trials. To do so, users can click the gear icon next to a parameter group and choose whether they want to use the "Trial Builder" interface or upload a .csv file with trial-by-trial values (Figure 6A).

**Figure 6. Advanced options to vary parameters across trials (A) and Trial Builder (B).**

The Trial Builder is a point-and-click, drag-and-drop interface that enables experimenters to define trial-by-trial sequences of parameter values using common experimental schedules (Figure 6B). First, users define the experimental phases (e.g., baseline, learning, washout). Second, for each phase, users select a sequence type:

- fixed sequence: parameter values are provided in order;
- randomized sequence: values are randomly sampled from a provided list (fully random with probability weights or blocked randomization);
- gradual ramp up/down: values change linearly between specified start and end points;
- statistical distribution: values are drawn from a uniform or normal distribution.

Third, users enter the required values in editable input fields – such as providing the list of values for fixed or random sequences, or providing the parameters of a distribution (e.g., mean and standard deviation).

Once the experiment design is complete, MovementVR Builder generates a .csv file with all the task parameters, including both constant and trial-varying values. Users can download this file using the Download button in the Builder interface and transfer it to the VR headset's internal memory (step-by-step instructions are provided on the website). This .csv file serves as the input for the MovementVR Application, so that all defined parameters are automatically applied when the task is executed.

**Step 2 with MovementVR Application**

The MovementVR Application consists of the main app on the VR headset that runs the motor task described in a previous section.

After step 1 is completed, the apk file for the MovementVR Application must be downloaded and the application installed on the VR headset device in order to conduct the experimental session. As mentioned, the input files created in step 1 should be transferred to the VR headset. Once these preparatory steps are completed, the experimental session can be initiated; the application will automatically integrate the specified paradigm and customizations. All instructions as well as the apk are provided on the website associated with MovementVR, in the MovementVR Application tab.

When using Meta Quest 2 or another VR headset with integrated cameras for hand tracking, the app is standalone. This means that no computer or external tracking device is needed – the data is automatically recorded upon executing the app. The website for MovementVR contains instructions for setting up alternative VR headsets that may not have integrated cameras, relying instead on the Leap Motion sensor for hand tracking.

**Step 3 with MovementVR Analyzer**

The MovementVR Analyzer exports data collected by the MovementVR Application. The data is automatically recorded upon execution and stored in the headset internal memory in binary (.bin) format. The binary files contain frame-by-frame 3d position of the headset and all the tracked joints in the hand (unless unselected by the user), as well as relevant task information (e.g., success or failure, time stamps for task events such as plate contact and trial completion). Data storage is limited by headset memory capacity, so users should ensure to offload the headset from time to time. The data can be exported by connecting the headset to a computer. The binary data can be processed in any software of choice. Alternatively, the MovementVR Analyzer provides a tool to convert the binary data to standard csv format – matching the format of common platform such as Vicon Nexus. To do this, the user would click the upload button on the MovementVR Analyzer webpage, select the binary files from the headset or computer memory, and then click the export button to download the converted csv file.

## **Discussion**

Virtual reality (VR) has emerged as a powerful tool for studying motor control and learning, offering an immersive and interactive environment while enabling precise experimental control (Arlati et al., 2022; Faria et al., 2023; Ingram and Wolpert, 2011; Levac et al., 2019). However, despite advancements in VR technology, its integration into movement research has remained limited due to technical challenges, lack of accessible tools, and the need for extensive programming expertise (Corti et al., 2022; Glegg and Levac, 2018; Haar et al., 2021; Levac et al., 2019; Ribeiro-Papa et al., 2016). MovementVR was developed to address these barriers by providing an open-source platform that facilitates the design and execution of motor tasks in VR without requiring coding expertise.

The platform was designed with two primary objectives: (1) improving the ecological validity of VR-based motor tasks and (2) ensuring accessibility for researchers and clinicians across disciplines. MovementVR incorporates physics-based object manipulation, real-time hand tracking, and a customizable task structure, allowing researchers to study motor learning under conditions that closely approximate real-world interactions. Unlike traditional VR paradigms that rely on simplified environments or rigid movement constraints, MovementVR enables naturalistic motor behaviors by eliminating the need for controllers and allowing direct hand-object interactions (Cesanek et al., 2024; Haar et al., 2021; Ingram and Wolpert, 2011; Levac et al., 2019). Moreover, by leveraging consumer-grade VR headsets with integrated hand tracking, MovementVR reduces financial and technical barriers, making high-quality VR-based motor research more accessible (Arlati et al., 2022; Levac et al., 2019).

The MovementVR platform was designed to study unconstrained naturalistic lifting motions and provides the ability to study gain adaptation and reinforcement learning. Nonetheless, our platform is not free of limitations. For example, our platform provides more flexibility than existing out-of-the-box VR tools, but to expand the utility of MovementVR to tasks beyond the lifting task and gain adaptation one would need to access and modify backend code and graphic application. For those who would like to do so, the Unity package from which the apk was created will be made available. Tracking is also limited to wrist and forearm; therefore, protocols that require upper arm tracking are not available. Finally, as with most VR applications, especially those relying on built-in cameras for hand tracking, optimal lighting conditions are necessary for best tracking, rendering, and recording of hand position.

MovementVR is freely available at https://movementvr.github.io/, and its documentation provides step-by-step guidance for researchers to design, implement, and analyze experiments. Future work will expand its capabilities with new motor learning paradigms, enhanced customization, and broader VR hardware compatibility. MovementVR represents a significant step toward making immersive and ecologically valid motor learning research more accessible, ultimately bridging the gap between experimental rigor and real-world applicability.

**Acknowledgements**

Supported by grant R35 NS122266 to AJB, T32 HD007414 to AJB, American Heart Association predoctoral 1407 fellowship 20PRE35180131 to CR.